\documentstyle[12pt]{article}
\newcommand{\norm}[1]{{\protect\normalsize{#1}}} \newcommand{\LAP}
{{\small E}\norm{N}{\large S}{\Large L}{\large A}\norm{P}{\small P}}

\newcommand{\be}{\begin{equation}}
\newcommand{\ee}{\end{equation}}
\newcommand{\bea}{\begin{eqnarray}}
\newcommand{\ena}{\end{eqnarray}}
\newcommand{\beano}{\begin{eqnarray*}}
\newcommand{\enano}{\end{eqnarray*}}
\newcommand{\sect}[1]{\setcounter{equation}{0}\section{#1}}
\newcommand{\vs}[1]{\rule[- #1 mm]{0mm}{#1 mm}} \newcommand{\hs}[1]{\hspace{#1
mm}}

\newcommand{\sg}[1]{\mbox{$\sigma_{#1}$}}

\newcommand{{\cg}}{\mbox{$\cal{G}$}}

\newcommand{\cl}{\mbox{$\cal{L}$}}

\newcommand{\prt}{\partial}

\newcommand{\mb}[1]{\hs{5}\mbox{#1}\hs{5}} 
\newcommand{\R}{\mbox{\hspace{.04mm}\rule{0.2mm}{2.8mm}\hspace{-1.5mm} R}}
\newcommand{\C}{\mbox{\hspace{1.24mm}\rule{0.2mm}{2.5mm}\hspace{-2.7mm} C}}

\newcommand{\Z}{\mbox{$Z\hspace{-2mm}Z$}} 
\newcommand{\N}{\mbox{\hspace{.04mm}\rule{0.2mm}{2.8mm}\hspace{-1.5mm} N}}

\newcommand{\CMP}[1]{Comm.\ Math.\ Phys.\ {\bf #1}}

\begin{document}
\renewcommand{\thefootnote}{\fnsymbol{footnote}} \newpage
\pagestyle{empty}
\setcounter{page}{0}
\rightline{September 93}

\vs{20}

\centerline{\LARGE{\bf Currents on Grassmann algebras}}

\vs{10}

\centerline{{\Large R. Coquereaux}}

{\it \centerline{Centre de Physique Th\'eorique} \centerline{CNRS Luminy - Case
907}
\centerline{F 13288 Marseille Cedex 9 (France)}}

\vs{4}

\centerline{{\Large E. Ragoucy}}

{\it \centerline{{\sl Laboratoire de Physique Th\'eorique} \LAP}
\centerline{B.P. 110}
\centerline{ F 74941 Annecy-le-Vieux Cedex (France)}}

\vs{10}

{\bf Abstract :}
We define currents on a Grassmann algebra $Gr(N)$ with $N$ generators as
distributions on its exterior algebra (using the
symmetric wedge product). We interpret the currents in terms of ${\Z}_2$-graded
Hochschild cohomology and closed currents in terms of cyclic
cocycles (they are particular multilinear forms on $Gr(N)$). An
explicit construction of the vector space of closed currents of
degree $p$ on $Gr(N)$ is given by using Berezin integration.

\vs{7}

{\bf Keywords :}

Grassmann algebras, super-algebras, supersymmetries, supermanifolds, Hochschild
and cyclic cohomologies, De Rham currents.

\vfill


\vs{4}

\begin{minipage}{4cm}
September 1993\\
CPT - 93 / P.2935
\end{minipage}
\hfill
\begin{minipage}{4cm}
\LAP - 440/93
\end{minipage}


\newpage
\pagestyle{plain}
\sect{Introduction}

\indent

Roughly speaking, De Rham currents are distributions that can be pictured as
possibly singular functions with several variables on a manifold. More
precisely they
extend the concept of distributions to the case of exterior forms. One can
generalize
currents to graded-commutative calculus, i.e., in the simplest case, to
Grassmann
algebras. The analogy with the classical situation is useful but can be, at
times,
confusing, since the space of exterior forms over a Grassmann algebra in
endowed with a
{\sl symmetric} wedge product in contradistinction with the usual exterior
algebra; in
particular the dimensionality of this space extends to infinity. With this
warning in mind
one notices that Grassmannian currents, like their classical counterpart in
commutative
geometry, i.e. the De Rham currents, and also like submanifolds may have
boundary or not.
The space of closed (i.e. without boundary) Grassmanian currents of degree $p$
over a
Grassmann algebra with $N$ generators is a finite dimensional ${\Z}_2$-graded
vector space.

\indent

In this paper, we show that these closed currents are in correspondence with
${\Z}_2$-graded cyclic cocycles over the Grassmann algebra: This is therefore a
${\Z}_2$-graded generalization of what is already
known in the case of manifolds \cite{AC1,AC2}. The situation is actually quite
simple because, in the case of Grassmann algebras, the super-commutative
analogue of De Rham cohomology is
trivial. In simple terms we shall determine the dimensionality of the space of
closed
currents and express them explicitly in terms of symmetric tensor (over the
Grassmann
algebra) whose Grassmann divergence is zero. The dimensionality of the space of
cyclic
cocycles (for general $p$) was already known after the work \cite{K} and an
explicit expression in the case $p=1$ was given for them was given in
\cite{CFRS}.

\indent

Our purpose here is\\
1) To interpret cyclic cocycles in term of closed Grassmannian currents\\ 2)
Build explicitly these objects in
terms of symmetric tensors over the Grassmann algebra, using the Berezin
integral.

\sect{Differential forms over Grassmann algebras and Grassmannian currents}

\subsection{Classical analogies}

\indent

{}From the algebraic point of view, it is convenient to consider the
Grassmann algebra $Gr(N)$ with $N$ generators $\theta^\mu, \mu \in
\{1,\ldots,N\}$ as the graded-commutative analogue of the commutative
algebra $A(N) \doteq C^\infty (\R^N)$ generated by the coordinates
$x^\mu$. Clearly, $\R^N$ (or better, $A(N)$ at the algebraic level) is the
simplest example of a
manifold. In the same way, $Gr(N)$ is the simplest example of a supermanifold.

In a course of elementary differential calculus, after having defined
integration over $\R^N$, one
builds the space $\Omega^p \doteq \Omega^p(A(N))$ of differential forms of
degree
$p$ and extend the definition of the integral to the whole of $\Omega\doteq
\oplus_p \Omega^p$. The space of differential forms over $\R^N$ is built as the
linear span (real coefficients) of products $dx^\mu \ldots dx^\rho$ with
$dx^\mu dx^\nu = - dx^\nu dx^\mu$. Then one proves a basic tool of
calculus, namely the Stokes' theorem that relates the integral of a
differential $d\omega$ of the form $\omega$ over a particular subspace $X$ to
the integral of the form $\omega$ over the
boundary $\partial X$ of $X$. Finally, one proves the Poincar\'e Lemma: On
$\R^N$, every
closed (i.e. $d\omega = 0$) $p$-form, $p>0$ is exact (i.e. $\exists \alpha \
with \ \omega =
d\alpha$). Thus, De Rham cohomology of $\R^N$ is trivial.

Physicists and mathematicians (also ingeneers!) know that, for many
applications, one has to use distributions rather than functions. Usual
distributions on $\R^N$ are therefore well known and are properly defined as
linear forms over an appropriate class of test functions. One can generalize
this concept of distributions to forms. Such distributions are usually called
De Rham currents, distributional forms or simply ``currents'' \cite{JD} and are
now a classical and necessary tool of analysis over $\R^N$ (or over manifolds).
In simple terms, a
current is a linear machine that produces a number out of a $p$-form. A simple
example of a
p-current $C$ can be obtained from any $p$-dimensional submanifold $S$ (more
properly from any
$p$-chain) by writing $\langle C, \omega \rangle \doteq \int_S \omega$.
Completely
antisymmetric contravariant tensors (so called ``$p$-vectors'') also define
currents. Indeed, let
${\cal L} \doteq Der (A(N))$ denote the space of derivations on $A(N)$;
elements of ${\cal L}$ are vector fields $\chi = \chi^\mu {\partial \over
\partial x^\mu}$. Let us call ${\cal L}^p$
the space of $p$-vectors, with elements $\chi = \chi^{\mu_1\mu_2\ldots\mu_p}
{\partial \over \partial x^{\mu_1}} \wedge {\partial \over\partial x^{\mu_2}}
\wedge {\partial \over \partial x^{\mu_p}}.$ As in the case of forms over
$A(N)$, this wedge product of $p$-vectors is
antisymmetric. Such a $p$-vector defines a $p$-current: Let $\omega =
\omega_{\mu_1\mu_2\ldots\mu_p} dx^{\mu_1} \wedge dx^{\mu_2} \wedge d x^{\mu_p}
\in \Omega^p$, one
defines the current $C_\chi$ associated to $\chi$ via the equation
\be
\langle C_\chi, \omega \rangle \doteq \int_{\R^N} \chi^{\mu_1\mu_2\ldots\mu_p}
\omega_{\mu_1\mu_2\ldots\mu_p} d^Nx
\ee
One can define a boundary operator $\partial$ mapping $p$-currents to $p-1$ as
follows
\be
\langle \partial C, \omega \rangle= \langle C, d\omega\rangle
\ee
Of course, currents can be
closed (i.e. $\partial C = 0$) or not.
If we only consider ``regular'' objects, we can also associate $p$-vectors to
currents and
since we shall only stay at a purely formal level, i.e. forget about problems
of analysis, this
is what we shall do in most of the following and therefore identify
$p$-currents and
$p$-vectors. If $\chi$ denotes a $p$-vectors with corresponding current $C$, we
shall call $\partial \chi$, with components ${\partial \over \partial
x^{\mu_1}}
\chi^{\mu_1\mu_2\ldots\mu_p}$ the $(p-1)$-vector associated with the current
$\partial C$. We shall often use the same notation $\chi$ or $C$ for both
$p$-vectors and currents.

In classical differential calculus (commutative geometry), the next step is to
generalize the
above notions to the case of manifolds. As everybody knows, a new phenomenon
appears: The
Poincar\'e lemma may fail (a form that is closed is no longer necessarily
exact!).
However, our aim, in the present article, is to study (some) properties of
Grassmann algebras
related to currents and not to study the corresponding properties on arbitrary
graded-commutative algebras. As stressed previously the former are simple
graded-commutative analogues of
$A(N)=C^\infty(\R^N)$: They are the simplest ``supermanifolds'' and, exactly as
for $\R^N$, De Rham cohomology is trivial. We can therefore stop here this
paragraph devoted to classical analogies.

Let us stress the fact that the space $\Omega$ of exterior forms over $\R^N$ is
isomorphic with $Gr(N)$ is certainly true but can be confusing in the context
of the present discussion. Let us repeat
again that the analogy that we use here is between $A(N)$ and $Gr(N)$. We now
turn our
attention to the Grassmanian analogue of $\Omega$, the algebra $\Lambda$ of
exterior forms over $Gr(N)$.

\subsection{Grassmann algebras (a compendium of basic results and notations)}

\indent

We summarize here some properties of Grassmann algebras and introduce a few
notations that we shall need in the sequel. More properties can be found in
\cite{B} or \cite{CJK}.

A Grassmann algebra $Gr(N)$ can be defined as a (real or complex) unital
algebra with $N$
generators $\theta^1,\theta^2\ldots,\theta^N$ satisfying the relations
$\theta^\mu \theta^\nu +
\theta^\nu \theta^\mu = 0$ and therefore $(\theta^\mu)^2 = 0.$ In the following
we choose the
basic field as the field $\C$ of complex numbers. As a vector space, it has
dimension $2^N$ and a possible basis is $\theta^I$ where $I$ is a multi-index
$I \in \{0, i, ij,\ldots, ij\ldots p,\ldots, 12\ldots N\}$ with $ 1\leq i < j <
\ldots < p\leq N$
together with the convention $\theta^I=1$ when $I=0$. $Gr(N) = {}^0Gr(N) +
{}^1Gr(N)$ is a graded vector space but also a graded commutative algebra:
Calling $\sg{a}$ the intrinsic grade of
$a$ (i.e. $0$ or $1$ on even or odd product of $\theta^\mu$ resp.), we have
$vu=(-1)^{\sg{u} \sg{v}}\ uv.$ It is sometimes useful to decompose $Gr(N)$ as
$Gr(N)= \C + {\cal N}$ where ${\cal N}$ denotes the space of nilpotents
elements of $Gr(N)$ and to notice that ${\cal N}$ is an ideal in $Gr(N)$. There
is a canonical homomorphism $\tau$
from $Gr(N)$ to $\C$, i.e., a character, defined by $\tau(1)=1$ and
$\tau|_{\cal N} = 0.$ An element
$u \in Gr(N)$ is invertible iff $\tau(u) \neq 0 $. In this case, calling
$\frac{u}{\tau(u)} = 1 + x$ one
gets $(1+x)^{-1} = 1 + \sum_{k=1}^N(-1)^kx^k$.

The generating system $\{\theta^\mu\}$ used to define $Gr(N)$ is by no means
unique; it is clear that one can make a change of generating system by using an
invertible matrix with scalar
coefficients but we should stress that one can also obtain new generating
systems for the algebra
$Gr(N)$ by choosing coefficients in $Gr(N)$ itself rather than in $\C$. More
precisely, one calls a
chart (or frame) on $Gr(N)$ a generating system $(\eta^\mu)_{\mu \in
\{1,\ldots, N\}}$ such that the $\eta^\mu$ are odd and such that
$\eta^1 \eta^2 \ldots \eta^N \neq 0$, one proves that these two conditions
imply $\eta^\mu
\eta^\nu + \eta^\nu \eta^\mu = 0.$ Let then $\{\theta\}$ be a frame of $Gr(N)$,
the corresponding Berezin integral is the element of the dual $Gr(N)^*$ given
by
\be
\int a d_\theta \doteq {\partial \over \partial \theta^N}{\partial \over
\partial
\theta^{N-1}}\ldots{\partial \over \partial \theta^1} a \ee
where $a \in Gr(N).$ In other words
this Berezin integral is equal to $a_{1,2,\ldots,N}$, the coefficient of $a$ on
the ``top
element'' $\theta^1\theta^2\ldots\theta^N$. We have to stress the fact that the
left hand side
of the previous equation is defined by the right hand side; in particular the
symbol $d_\theta$ has nothing to do with a 1-form $d\theta$: see section
\ref{sec.2.3}.
If $\{\eta\}$ is
another chart of $Gr(N)$ --that we obtain from $\{\theta\}$ by an invertible
matrix with coefficients in ${}^0Gr(N)$, the corresponding Berezin integral is
defined in the same way but we have the relation
\be
\int d_\eta \ a = \int d_\theta \ \left|{\partial \theta \over \partial
\eta}\right| \times a
\ee
where
\be
\left|{\partial \theta \over \partial \eta}\right| \doteq Det \ \left[{\partial
\theta^\mu \over \partial \eta^\nu}\right] \in Gr(N) \ee
The fact that both Berezin integrals associated with charts $\theta$ and $\eta$
are related in
this way illustrates another interesting property of Grassmann algebras, namely
the fact
that the dual $Gr(N)^*$ of $Gr(N)$ is simply generated as a right module over
$Gr(N)$, in other
words, if $\phi \in Gr(N)^*$ is such that
$\phi(\theta^1\theta^2\ldots\theta^N)\neq0$ then $\phi$ generates $Gr(N)^*$,
i.e. $\phi \ Gr(N) = Gr(N)^*.$

\subsection{Derivations and differential forms over $Gr(N)$\label{sec.2.3}}

\indent

We call ${\cal L} \doteq Der(Gr(N))$ the space of (graded) derivations on
$Gr(N)$, i.e. the
analogue of the space of vector fields on $\R^N$. Calling $\theta =
\{\theta^1,\ldots,\theta^N\}$
a chart in $Gr(N)$ we notice that ${\cal L}$ is a left $Gr(N)$-module (actually
a free module) and that $\{ {\partial \over \partial \theta^1},\ldots,{\partial
\over \partial \theta^N}\}$ is a
basis of the module ${\cal L}.$ In other word, an arbitrary element $\xi$ of
${\cal L}$ can be
written $\xi = a^\mu{\partial \over \partial \theta^\mu}$ with $a^\mu \in
Gr(N).$
As a vector space, ${\cal L}$ has a basis $\{\theta^I {\partial \over
\prt\theta^\mu}\}$ where $I$ is a multi-index (with $2^N$ possible values) and
$\mu
\in \{1,2\ldots,N\}$ and it is also a graded vector space: ${\cal L} =
{}^0{\cal L} \oplus {}^1{\cal L}$, with
$\sg{(\theta^I\frac{\prt}{\prt\theta^\mu})} \doteq card(I)+1\ \ mod \, 2$.
Notice that, in particular, $\frac{\prt}{\prt\theta^\mu}$ is odd.

In the classical (commutative) case the space of vector fields is not only a
$A(N)$-module but also a Lie algebra. Here the situation is
analogous: ${\cal L}$ is a Lie superalgebra for the bracket $[\xi,\eta]=\xi\eta
-
(-1)^{\sg{\xi} \sg{\eta}}\, \eta \xi $ and, in particular
$\frac{\partial}{\prt\theta^\mu}\frac{\partial}{\prt\theta^\nu}+
\frac{\partial}{\prt\theta^\mu}\frac{\partial}{\prt\theta^\mu}=0.$

We call ${\cal L}^*$ the dual of ${\cal L}$ on $Gr(N)$ and denote by
$d\theta^\mu$ the elements of
${\cal L}^*$ defined by $<d\theta^\mu,\xi>=(-1)^{\sg{\xi}} \xi(\theta^\mu),$ in
particular
$<d\theta^\mu,{\partial \over \partial \theta^\nu}> = - \delta^\mu_\nu.$ An
arbitrary element $\lambda$ of $\cl^*$ is an arbitrary linear combination of
the $\theta^Id\theta^\mu$. For an arbitrary element
$a\in Gr(N)$, one defines the $d$ operator as $da= d\theta^\mu {\partial \over
\partial \theta^\mu}a\,
\in {\cal L}^*.$ Notice that ${\cal L}^*$ is also a left $Gr(N)$-module. ${\cal
L}^*$ is also a graded vector space and we define
$\sg{(\theta^I d \theta^\mu)} \doteq card(I) \, \, mod \, 2$. Notice that this
definition of the grading makes the $d\theta$'s even as we shall discuss it
now.

We now call $\Lambda^k \doteq \Lambda^k({\cal L},Gr(N))$ the space of
$Gr(N)$-valued, graded alternate $Gr(N)$-linear $k$-forms over ${\cal L}.$
Notice that $\Lambda^1 = {\cal L}^*$.
Elements of $\Lambda \doteq \sum_{0}^{\infty} \Lambda^k$ can be multiplied and
$\Lambda$
becomes the Grassmann analogue of the usual algebra of exterior forms in
$\R^N$. This space can also be considered as the exterior algebra over ${\cal
L}^*$
{\sl but} the wedge product between $1$-forms is now ${\sl symmetric}$.
Actually this is quite
natural: $1$-forms anti-commute in commutative geometry but they commute in
anti-commutative geometry!
In other words (and in practice) it is enough to remember that $d$ is odd, as
well as the $\theta$'s so that $d\theta$ is even and we can use the usual
Milnor'sign rule. To avoid confusion, we denote this product `$\vee$' in the
sequel, and write for instance $d\theta^\mu\vee d\theta^\nu= d\theta^\nu\vee
d\theta^\mu$
There is a only one $d$ operator
of square $0$ that extends the $d$ of the one-forms to the whole of $\Lambda$
and this space becomes then a $\Z_2$-graded and bigraded differential algebra
(the bigrading is with respect to $\N$). Indeed, an arbitrary element of
$\Lambda^k$ can be written $\lambda =
\lambda_{\mu_1\mu_2\ldots\mu_k}d\theta^{\mu_1}\vee\ldots \vee d\theta^{\mu_k}$
where
$\lambda_{\mu_1\mu_2\ldots\mu_k}$ is a covariant (down indices) symmetric
tensor with coefficients in $Gr(N)$, therefore $\lambda_{\mu_1\mu_2\ldots\mu_k}
= \sum_I\lambda_{\mu_1\mu_2\ldots\mu_k}^I \theta^I$ where $\theta^I$ is a multi
index. Each single term of this last sum is of the kind
$\lambda_{\mu_1\mu_2\ldots\mu_k}^{i_1\ldots i_p}
\theta^{i_1}\ldots\theta^{i_p}$ and this is why we have an $\N$-bigrading
$(k,p)$. Because of the commutativity between the $d\theta$ themselves, graded
commutativity of the wedge product depends only the $\Z_2$ grading $p \, mod \,
2$. One can also introduce an interior product and a Lie derivative in
$\Lambda$ but we shall not use those concepts and we refer to \cite{KJ} and
\cite{CJK} for more details. What is important for us is that $d= d\theta^\mu
{\partial \over \partial \theta^\mu}$; therefore, the above notation we can
write $\omega\doteq d\lambda$ with \be
\omega= \frac{1}{(k+1)!} \omega_{\mu_0\mu_1\dots\mu_k} d\theta^{\mu_0}
\vee\dots\vee d\theta^{\mu_k}
\ee
and
\be
\omega_{\mu_0\mu_1\dots\mu_k} \doteq \prt_{[\mu_0}
\lambda_{\mu_1\mu_2\dots\mu_k]}
\ee
where the symbols $[\dots]$ denote complete symmetrization and where \be
\prt_{\mu_0}\lambda_{\mu_1\mu_2\ldots\mu_k} = \sum_{i_1,\ldots,
i_p}\sum_{s=i_1}^{i_p} (-1)^{(s+1)} \lambda_{\mu_1 \ldots\mu_k}^{i_1\ldots
i_s\ldots i_p} \delta^{i_s}_{\mu_0} \theta^{i_1}\ldots \hat{\theta^{i_s}}\ldots
\theta^{i_p}
\ee
As usual, the hat on $\theta^{i_s}$ means that this factor has been omitted in
the product. Notice that because of the symmetry properties of the $\vee$
product, the family $d\theta^{\mu_1}\vee\dots\vee d\theta^{\mu_k}$ is a
generating family for vectors, but is not a basis of $\Lambda^k$, unless we
impose an ordering constraint like $\mu_1\leq\mu_2\leq\dots\leq\mu_k$ (this is
analogous to the situation prevailing in the case of usual exterior forms).

At the dual level, currents -- that we take as $p$-vectors -- will also be
represented by
{\sl contravariant (up-indices) symmetric tensors with coefficients in}
$Gr(N),$ i.e. $\phi =
\phi^{\mu_1 \ldots \mu_k} \partial_{\mu_1}\vee\ldots\vee \partial_{\mu_k}.$ The
pairing between
$p$-vectors and $p$-forms is given by
\be
\langle \phi, \lambda \rangle \doteq (-)^k \int d^N_\theta \phi^{\mu_1 \ldots
\mu_k} \lambda_{\mu_1 \ldots \mu_k} \ee

A well known result of elementary tensorial calculus (or of quantum mechanics
of systems with identical bosonic particles) tell us that the space of
symmetric tensors of order $p$ built on a vector space of dimension $N$ has
dimension
$C_{N+p-1}^p$, but in the present case, the coefficients belong to the
Grassmann algebra $Gr(N)$, itself of dimension $2^N$, therefore the space of
differential forms over $Gr(N)$ has dimension $dim (\Lambda^p({\cal
L},Gr(N))=2^N C_{N+p-1}^p.$ Notice that, in contrast
with the case of usual exterior forms over ${\R}^N$ or over a finite
dimensional manifold, the space of all such forms (direct sum of all spaces
$\Lambda^p$ for all values of $p$) is infinite dimensional since
$dim(\Lambda^p)$ grows with $p$.

The pairing between contravariant and covariant symmetric $Gr(N)$-valued
tensors (i.e. currents and forms) enables one to define a boundary operator
$\partial$ already mentioned before. Writing $\langle \partial \psi, \lambda
\rangle \doteq \langle \psi, d\lambda \rangle$ for a
$p$-vector $\psi$ and a $(p-1)$-form $\lambda$ leads to \be
\prt\psi^{\mu_1\dots\mu_{p-1}} = \sum_{j=1}^N \partial_j \tilde \psi^{j
\mu_1\ldots\mu_{p-1}}
\ee
where ``\ $\tilde {}$ '' denotes the involution of $Gr(N)$ associated with the
${\Z}_2$
grading, i.e. a change of sign in front of odd monomials (for instance if $a =
2 \theta^3 +
5\theta^1 \theta^2 - \theta^1 \theta^2 \theta^3$ then $\tilde a = -2 \theta^3 +
5\theta^1
\theta^2 + \theta^1 \theta^2 \theta^3$).

Notice that a $Gr(N)$-valued contravariant tensor $\phi^{\mu_1 \mu_2 \ldots
\mu_p}$
defines a $Gr(N)$-valued differential operator of degree $p$, namely
$P_\phi \doteq \phi^{\mu_1 \mu_2 \ldots \mu_p}
\partial_{\mu_1}\partial_{\mu_2}\ldots\partial_{\mu_p}$ where the product of
symbols is now just the composition law
$\partial_{\mu_1}\partial_{\mu_2}\ldots\partial_{\mu_p}$ \underline{but} as
$\prt_\mu\prt_\nu=- \prt_\nu\prt_\mu$, $P_\phi$ is identically zero when $\phi$
is symmetric. In what follow, we will use $p$-vectors
$\phi=\phi^{\mu_1\ldots\mu_p} \prt_{\mu_1}\vee\ldots \vee\prt_{\mu_p}$ with a
symmetric wedge product $\vee$. Differential operators and $p$-vector coincide
{\sl only} when $p=1$. In that case, the quantity $\partial \phi$, which is
the boundary of $\phi,$ can also be
called the (graded) divergence of the operator $P_\phi$ with, of course,
$\partial \phi =
Div(P_\phi).$ This identification is used, for instance in \cite{CFRS}.

The last thing to study in this paragraph is the Grassmann analogue of De Rham
cohomology of
differential forms (or Grassmann homology of currents). Let us denote by
$Z^p_{DR} \doteq \{\omega \in \Lambda^p, d\omega=0\}$, $B^p_{DR} \doteq
\{\omega \in \Lambda^p, \omega=d\psi\}$ and $H^p_{DR} \doteq Z^p_{DR}/
B^p_{DR}.$ One can prove that this cohomology is trivial: $H_{DR}=0.$ Indeed
one can construct (see for instance \cite{CJK}) an operator $h$ from
$\Lambda^p$ to
$\Lambda^{p-1}$ such that $d h + h d = 1$, therefore, given a closed form
$\omega$, one defines $\psi = h\omega$ and shows that $\omega$ is exact
($\omega = d\psi$) by using the previous relation between $d$ and $h$. From
this point of view, Grassmann algebras are topologically
trivial (as their ``commutative''counterpart): The Poincar\'e lemma holds.

Before ending this paragraph, let us remark that the symbol $d_\theta$
appearing in the Berezin integral should \underbar {not} be considered as a
shorthand for
$d\theta^1\vee d\theta^2\ldots \vee d\theta^N$, at least not with the symmetric
product
$\vee$, as defined above, in mind. Indeed the law of change of variables in the
Berezin
integral shows that this integral should be transformed by a global $-1$
multiplicative constant
upon a parity-odd change of chart such that \be
\{\theta^1, \theta^2, \theta^3, \ldots, \theta^N\} \longrightarrow \{\theta^2,
\theta^1,
\theta^3, \ldots ,\theta^N \}
\ee
and this is clearly incompatible with a symmetric wedge product.
One possibility is to introduce a new (antisymmetric) product -- that one would
call ``$.$''
rather than ``$\vee$''-- between the $d\theta$'s, just for the sake of making
the notation $\int f(\theta) \ d\theta$ compatible with the law of change of
variables. Another possibility is to
refrain from writing a Grassmanian volume element in Berezin integrals, but
this notation is now quite popular $\ldots$ Let us only remember that the
``natural'' product between $d\theta$'s is the symmetric one and that it is
that one that we have been using and shall use in the following, even if we do
not write explicitly the wedge symbol. Actually, Berezin himself, in his book
\cite{B} does not use the symbols $d \theta^i$ in his
integrals but the symbol $d_{\theta^i}$ and warns the reader (footnote page 76)
that he should \underline{not} confuse the $d_{\theta^i}$ with the differential
$d{\theta^i}$: it is his notation that we have choosen, and we will write for
the volume element:
\be
d_\theta^N = d_{\theta^1}d_{\theta^2}\ldots d_{\theta^N} \ee

\sect{Hochschild and cyclic cohomologies for Grassmann algebras\label{sec2}}

\subsection{Hochschild cohomology}

\indent

The definition of Hochschild cohomology of an arbitrary non-commutative algebra
$A$ with values in the bimodule $A^*$ --its dual-- can be found, for instance
in \cite{AC1}.
The Hochschild $p$-cochains can be taken, in the present case, as arbitrary
complex-valued
$(p+1)$-linear forms over $Gr(N).$ Their space will be denoted by ${\cal C}^p.$
Notice that this space is naturally ${\Z}_2$-graded (${\cal C}^p = {}^0{\cal
C}^p \oplus {}^1{\cal C}^p$)
and that $dim \ {\cal C}^p = (2^N)^{p+1}=2^{N(p+1)}$ with $dim \ {}^0{\cal
C}^p= dim \ {}^1{\cal
C}^p =2^{N(p+1)-1} .$ The generalization of Hochschild cohomology to
${\Z}_2$-graded
algebras can be found, for instance in \cite{DK} and the corresponding
coboundary operator $b$ mapping ${\cal C}^p$ to ${\cal C}^{p+1}$ and satisfying
$b^2=0$ is defined as follows:
\bea
[b\Phi](a_0,a_1,\ldots,a_{p},a_{p+1}) &\doteq& \sum_{j=0}^p (-1)^j
\Phi(a_0,\ldots,a_ja_{j+1},\ldots,a_{p+1}) \nonumber\\ &&+
(-1)^{p+1}(-1)^{\epsilon_{p+1}}
\Phi(a_{p+1}a_0,a_1,\ldots,a_p)
\ena
with
\be
\epsilon_{p+1} = (\sg{a_{p+1}}) \times \sum_{i=0}^p \sg{a_i}
\ee
The space of Hochschild cocycles and coboundaries will be denoted by $Z^p$ and
$B^p$ with
cohomology $H^p \doteq Z^p / B^p.$

We want to show that, in the case of Grassmann algebras, Hochschild cocycles
are in one
to one correspondence with $p$-currents. We choose therefore \be
\phi = \phi^{\mu_1\ldots\mu_p} {\partial \over \partial
\theta^{\mu_1}}\vee{\partial \over \partial
\theta^{\mu_2}}\vee\ldots\vee{\partial \over \partial \theta^{\mu_p}}
\ee
where
$\phi^{\mu_1\ldots\mu_p} \in Gr(N)$ is a {\sl symmetric} tensor (remember that
$\vee$ is symmetric). The corresponding Hochschild cochain $\Phi$ is defined by
\bea
\Phi(a_0,a_1,\ldots,a_p)
&=& \langle \phi,a_0da_1\ldots da_p\rangle \nonumber\\ &=& (-)^\epsilon \int
d^N_\theta \phi^{\mu_1\ldots\mu_p} \hskip 10pt a_0 (\partial_{\mu_1}a_1)
(\partial_{\mu_2}a_2) \ldots (\partial_{\mu_p}a_p) \\ \mb{with} \epsilon &=&
\frac{p(p+1)}{2} +\sum_{i=0}^{p} (p-i)\sg{a_i} \nonumber
\ena
It is a
simple exercise to show that $b\Phi =
0$, hence $\Phi$ is a Hochschild cocycle: $\Phi \in Z^p.$

Conversely, let $\Phi$ an arbitrary $p$-Hochschild cocycle. It has, {\it a
priori}, no reason to be symmetric. We have to symmetrize it and define a $p$
current as:
\be
\langle \phi, a_0da_1\ldots da_p\rangle \doteq \sum_{s \in {\cal S}}
\Phi(a_0,a_{s(1)},\ldots,a_{s(p)}) \label{eq.sym} \ee
Starting with some Hochschild cochain $\Psi$, one can show that if
$\Phi=b\Psi$, then the current $\phi$ corresponding to this Hochschild
coboundary $\Phi$ is zero. Indeed, using the definition of $b$ and, after the
symmetrization process described in eq. (\ref{eq.sym}), all the terms vanish
pair-wise.
This implies that we can identify
the space of Grassmannian currents (elements of $(\Lambda^p)^{*}$) with
Hochschild cohomology $H^p$. As a consequence, $dim(H^p) =
2^N C_{N+p-1}^p.$ Notice that
identification of De Rham currents with Hochschild cocycles seems to be a quite
general
property, since it also holds in the classical case of manifolds \cite{AC1}
(with the difference, of course, that the dimensionality of $H^p$ grows with
$p$ in the Grassmanian case whereas it is
zero as soon as $p$ exceeds the dimension of the manifold in the classical
--commutative-- case).
We do not study here the structure of general supermanifolds --Grassmann
algebra being the simplest example-- but we conjecture that, modulo an
appropriate definition of currents
generalizing the one given previously, the same result should hold.

Notice that a Grassmann algebra is both \Z-graded and $\Z_2$-graded; our
previous
construction of the correspondance between currents and Hochschild cocycles
uses implicitely the underlying \Z-grading. Indeed, it uses the Berezin
integral, but the value of this integral will change if we make a change of
chart with coefficients in ${}^0Gr(N)$ rather than in \C. Such a change a chart
would preserve
the $\Z_2$-grading but modify the \Z-grading.

Finally, let us remark that it may help the reader to think of $Gr(N)$ as the
fermionic Fock space
built over a fermionic system with $N$ levels (i.e. a complex vector space of
dimension $N$) and of Hochschild $p$-cocycles as bosonic transitions
associating a number to $(p+1)$ elements of the fermionic Fock space
(transitions are of bosonic type because of the symmetry properties of the
cocycles).

\subsection{Cyclic cohomology (results)\label{sec3.2}}

\indent

The cyclicity operator $\lambda$, in the ${\Z}_2$-graded case, is defined by
\be
[\lambda \Phi](a_0,a_1,\ldots,a_p) \doteq (-1)^p(- 1)^{\epsilon_p}
\Phi(a_p,a_0,\ldots,a_{p-1}) \ee
where
\be
\epsilon_p \doteq (\sg{ a_p})\times
\sum_{i=0}^{p-1}\sg{ a_i}
\ee
and $\Phi$ is a cyclic cochain iff $\Phi o \lambda = \Phi.$ The cyclic complex
\cite{AC1} is a
subcomplex of Hochschild complex.
In other words, cyclic cochains of degree $p$ on the Grassmann algebra are
represented by $(p+1)$-linear forms on $Gr(N)$ --i.e. Hochschild
cochains -- which are cyclic; their space will be denoted ${\cal C}_\lambda^p.$
A cyclic cochain
is therefore characterized by the value that it takes on $(p+1)$ basis elements
chosen among the
$2^N$ that span $Gr(N).$ Cyclic invariance shows therefore that $dim \ {\cal
C}_\lambda^p =
2^{N(p+1)}/(p+1).$ Again this vector space is graded ${\cal C}_\lambda^p =
{}^0{\cal C}_\lambda^p
\oplus {}^1{\cal C}_\lambda^p$ and $dim \ {}^0{\cal C}_\lambda^p = dim \
{}^1{\cal C}_\lambda^p = 2^{N(p+1)-1}/(p+1).$
The space of cyclic cocycles and coboundaries will be denoted by $Z_\lambda^p$
and $B_\lambda^p$ with cohomology $H_\lambda^p \doteq Z_\lambda^p /
B_\lambda^p.$
In the case of manifolds, it can be shown \cite{AC1} that $H_\lambda^k=Ker \
\partial_k \oplus H^{k-2}_{DR} \oplus H^{k-4}_{DR} \oplus \ldots, $ where
$H^k_{DR}$ denotes the homology of De
Rham currents (or of differential forms, at the dual level). When De Rham
(co)homology is
trivial one just gets $H_\lambda^p=Ker \ \partial_p $ where $\partial_p$ is the
boundary operator
acting on currents. This result suggests, by analogy with the
classical case, that one should find something similar in the
Grassmanian case, i.e. we expect $H_\lambda^p$ to be essentially given by $Ker
\partial_p$, where
$\partial_p$ is the boundary operator acting on Grassmanian currents, as
defined in section \ref{sec2}. By ``essentially'', we mean precisely the
following: \be
H_\lambda^p (Gr(N)) = Ker \ \partial_p \oplus H_\lambda^p({\C}) \ee
where $H_\lambda^p({\C})$ is equal to ${\C}$ if $p$ is even and $0$ if $p$ is
odd.
This is what we show in the following. More precisely, we obtain the following
results:

\indent

{$\bullet$}
Closed currents can be expressed as symmetric multilinear forms on $Gr(N)$
which are of vanishing graded divergence (i.e. $\prt\psi=0$).

\indent

{$\bullet$}
By direct counting, the space of such currents of degree $p$ is a graded vector
space
\be
Ker\prt_p\doteq V^p(N) = {}^0V^p(N) \oplus {}^1V^p(N) \ee
of dimensions $dim \
{}^0V^p(N) = 2^{N-1}
\gamma^N_p - (-1)^p$ and $dim \ {}^1V^p(N) = 2^{N-1} \gamma^N_p$ with
\bea
\gamma^N_p &=& \sum_{i=1}^{p+1}(-1)^{(i-1)} C_{N+p-i}^{p-(i- 1)} \nonumber\\
&=& C_{N+p-1}^p - C_{N+p-2}^{p-1} + C_{N+p-3}^{p-2} - \ldots + (-)^p C^0_{N-1}
\ena

\indent

{$\bullet$}
To any closed current, one can associate a cyclic cocycle.

\indent

{$\bullet$}
When $p$ is even, there will be a (unique, up to scalar multiple) canonical
cyclic cocycle of
degree $p$ that we shall call $\tau$. This cocycle cannot be expressed in terms
of closed
currents.

\indent

{$\bullet$}
To any cyclic cocycle which has no component along $\tau$ one can associate a
closed current.

An explicit basis of $H_\lambda^p$ can therefore be found by using the previous
results along with
Berezin integration.
\smallskip
The cyclic cohomology groups are therefore $H_\lambda^p(Gr(N)) =
H_\lambda^p({\C})
\oplus V^p(N)$ where $H_\lambda^p({\C})$ denotes the cyclic cohomology of
complex numbers (it is
known that $H_\lambda^{2s}({\C}) = {\C}$ and $H_\lambda^{2s+1}({\C}) = 0$) and
where $ V^p(N)$ is the graded vector space introduced previously. The
dimensionality of $H_\lambda^p(Gr(N))$ has been calculated in the literature by
using K\"unneth-like formulae \cite{K}. The novel
feature here is the relation of cyclic cocycles to Grassmanian currents and
their explicit
expression in terms of $Gr(N)$-valued symmetric multilinear forms with
vanishing graded
divergence; as a by-product, we recover the known result concerning the
dimension of
$H_\lambda^p(Gr(N))$.

Notice that the dimensionality of the space of non-trivial cyclic cocycles is
given (up to a
global factor $2^N$ and up to $1$ when $p$ is even) by the alternated partial
sums along NW-SE
diagonals of the Pascal triangle.

\subsection{Closed currents, divergence-free Grassmann- valued contravariant
symmetric tensors and cyclic cocycles}

\indent

We already showed that a closed current of degree $p$ (or a closed $p$-vectors)
can be
written in terms of a divergence-free differential operator of degree $p$, i.e.
in terms of a
Grassmann valued contravariant symmetric tensor $\phi^{\mu_1\mu_2\ldots\mu_p}$
satisfying $\sum_{j=1}^N \partial_j \tilde \phi^{j \mu_1\ldots\mu_p} = 0.$ It
remains to show that it defines a cyclic cocycle $\Phi$. We already know
(because it is a current) that it defines a Hochschild
cocycle. The only thing to check is that it is cyclic. The following sequence
of equalities establishes the claimed property.
\bea
\Phi(a_0,a_1,\ldots,a_p) &=& \int d^N_\theta\ \phi^{\mu_1\ldots\mu_{p-1}\mu_p}
\
a_0(\partial_{\mu_1}a_1) \ldots(\partial_{\mu_{p-1}}a_{p-1})
(\partial_{\mu_p}a_p) \nonumber\\
&=& (-1)^p(-1)^{\epsilon_p} \int
d^N_\theta\ \phi^{\mu_1\ldots\mu_{p-1}\mu_p} \ (\partial_{\mu_p}a_p)
a_0(\partial_{\mu_1}a_1) \ldots(\partial_{\mu_{p-1}}a_{p-1}) \nonumber\\
&=& + (-1)^p(-1)^{\epsilon_p} \int
d^N_\theta\ \phi^{\mu_1\mu_2\ldots\mu_{p-1}\mu_p}\ a_p
(\partial_{\mu_p}a_0)(\partial_{\mu_1}a_1) \ldots(\partial_{\mu_{p-1}}a_{p-1})
\nonumber\\
&=& (-1)^p(-1)^{\epsilon_p} \int
d^N_\theta\ \phi^{\mu_2\mu_3\ldots\mu_p\mu_1}\ a_p
(\partial_{\mu_1}a_0)(\partial_{\mu_2}a_1) \ldots(\partial_{\mu_p}a_{p-1})
\nonumber\\
&=& (-1)^p(-1)^{\epsilon_p} \int
d^N_\theta\ \phi^{\mu_1\mu_2\ldots\mu_{p-1}\mu_p}\ a_p
(\partial_{\mu_1}a_0)(\partial_{\mu_2}a_1) \ldots(\partial_{\mu_p}a_{p-1})
\nonumber\\ &=& [\lambda \Phi](a_0,a_1,\ldots,a_p]
\ena
Where the symbol $\epsilon_p$ has the same meaning as before. In the above
calculation, we use, in turn the following properties
\indent

{1.} The fact that a current $\phi$ defines a Hochschild cocycle $\Phi.$\\
\indent
{2.} (Graded) Commutation of the element $\partial_{\mu_p}a_p$ with the other
elements.\\ \indent
{3.} Integration by part (with a `` $+$ '' sign) and the fact that $\phi$ is
closed, i.e.
$\partial \phi = 0.$\\
\indent
{4.} Relabelling of indices.\\
\indent
{5.} The fact that $\phi^{\mu_1\ldots\mu_p}$ is symmetric.\\
\indent
{6.} Definition of the cyclicity operator $\lambda.$

\indent

Notice that the Hochschild
property $b\Phi = 0$ is translated in
the fact that $\Phi$ can be written as a current $\phi$, whereas the cyclicity
property $\lambda
\Phi = \Phi$ imposes that this current is closed $\partial \phi = 0$.
Actually, it is exactly the same thing in the classical case of commutative
calculus.

Notice also that we did not prove that {\sl all} cyclic cocycles (or cyclic
cohomology classes)
can be expressed in terms of non-zero closed currents; indeed it is possible to
exhibit
a non-trivial cyclic cocycle which is trivial as a Hochschild cocycle (i.e. is
a
Hochschild coboundary). In such a case the corresponding current is zero. This
case actually
happens only when $p$ is even and the situation is easy to control since there
is only one such
cyclic cohomology class that we shall call $\tau.$

\subsection{The hierarchy of cocycles $\tau$\label{sec3.4}}

\indent

Let $\theta = \{\theta^\mu\}_{\mu \in \{1,\ldots,N\}}$ be a chart in $Gr(N)$
and
$\{\theta^I\}_{I\in\{0,\ldots,2^N\}}$ denote the corresponding lexicographic
basis. We will denote by $\epsilon_I$ the dual basis of $Gr(N)^*$, i.e.
$\epsilon_I(\theta^J)=\delta_I^J$.
Clearly the direct product $(Gr(N))^{\times p}$ is spanned by
$\epsilon_{I_1,I_2,\ldots,I_p},$
where
\be
\epsilon_{I_1,I_2,\ldots,I_p}(\theta^{J_1},
\theta^{J_2},\ldots,\theta^{J_p})=\delta_{I_1}^{J_1}\delta_{
I_2}^{J_2}\ldots\delta_{I_p}^{J_p}.\label{eq.eps} \ee
In section \ref{sec2} we already called $\tau \doteq \epsilon_0$ the
fundamental character of $Gr(N)$,
i.e. $\tau$ assigns to each element of the Grassman algebra its ``scalar''
part. We keep the same notation for
\be
\tau \doteq \epsilon_{0,0,\ldots,0}
\ee
 From its definition, we see that, considered as a multilinear form on $Gr(N)$,
$\tau$ vanishes on homogeneous elements unless all elements $a_i$ are
${\C}$-numbers. In this case,
$\tau(a_0,a_1,\ldots,a_p)$ is just equal to the product $a_0 a_1\ldots a_p$.
{}From a direct computation, it is clear that $b\tau = 0$ if and only if $p$ is
even. For example
\be
[b\tau](a_0,a_1,a_2,a_3)=(a_0a_1)a_2a_3 -
a_0(a_1a_2)a_3+a_0a_1(a_2a_3)-(a_3a_0)a_1a_2 = 0 \ee
When $p$ is even, $\tau$ is therefore a Hochschild cocycle, moreover, $\tau$ is
also clearly cyclic. It is the b of something (Hochschild cochain) but this
Hochschild cochain is not
cyclic. Therefore, $\tau$ is a non trivial cyclic cocycle. However, if we try
to associate a current to $\tau$, we discover that the corresponding current
is strictly zero. What we just computed is nothing else than the cyclic
cohomology of complex
numbers: it is trivial when $p$ is odd and one- dimensional when $p$ is even.
What happens, as illustrated above, is that this hierarchy of cocycles is also
part of the cyclic cohomology of
Grassmann algebras. For us, this is the uninteresting part and we shall single
it out in the
sequel.

\subsection{The dimension of the space of closed currents of degree $p$ on
$Gr(N)$}

\indent

For big values of $N$ and $p$, the list of algebraic constraints imposed by the
equations $b\Phi = 0$ and $\lambda \Phi = \Phi$ becomes rather large
($2^{N(p+1)} + 2^{Np}$ equations) and a direct attempt of solving these
equations would become a formidable task! Fortunately, the fact that we were
able to recast these constraints in terms of divergence equations for symmetric
tensors will help us.
We already gave the general formulae for $dim V^p(N)$ in section \ref{sec3.2}
--with $V^p(N) \doteq Ker \partial_p$. We now indicate two different proofs of
those results.

\indent

{\sl First method}

\indent

The first method rests on an elementary induction proof and on the observation
that there exists a vector space isomorphism between $H^p$ -- that we identify
here with the space of
$Gr(N)$-valued symmetric tensors of order $p$ -- and the vector space $V^p(N)
\oplus V^{p-1}(N).$ Indeed, let $\phi^{\mu_1\ldots\mu_p}$ a totally symmetric
tensor. There are two possibilities: Either
$\partial_{\mu_1}\phi^{\mu_1\ldots\mu_p} = 0$ or
$\partial_{\mu_1}\phi^{\mu_1\ldots\mu_p} = \psi^{\mu_2\ldots\mu_p} \neq 0.$ In
the first case, we have $\phi \in V^p(N).$ In the second
case, $\partial_{\mu_2}\psi^{\mu_2\ldots\mu_p} =
-\partial_{\mu_1}\partial_{\mu_2}\phi^{\mu_1\ldots\mu_p}=0$ since
$\partial_{\mu_2}\partial_{\mu_1}=-
\partial_{\mu_1}\partial_{\mu_2}$ and
$\phi^{\mu_1\mu_2\ldots\mu_p}=\phi^{\mu_2\mu_1\ldots\mu_p}.$ Thus $\psi \in
V^{p-1}.$ Therefore, we have constructed a linear map from $H^p$ onto $V^{p-1}$
with kernel $V^p$. This establishes the existence of the above vector space
isomorphism.
Consequently $dim \ H^p = dim \
V^p + dim \ V^{p-1}.$ The induction goes as follows: When $p=0$,
the space $H_\lambda^0$ is just the dual of $Gr(N)$ -- the conditions $b\Phi=0$
and $\lambda \Phi = \Phi$ are trivially satisfied-- (actually ${\cal
C}^0=Z^0={\cal C}_\lambda^0=H_\lambda^0=Gr(N)^*$.) The dual is, as $Gr(N)$
itself, of dimension $2^N$, and therefore $dim \ V^0 = 2^N - 1$ (since we
single out the character $\tau$). We already know the dimension of $H^p$,
namely $2^N \
C_{N+p-1}^p$. Therefore
\bea
dim \ V^p &=&2^N \left\{
C_{N+p-1}^p-C_{N+p-2}^{p-1}+\ldots+(-1)^{p-1} C_N^1\right\} -(-1)^{p-1}(2^N-1)
\nonumber\\
&=& 2^N\left\{C_{N+p-1}^p - C_{N+p-2}^{p-1} + C_{N+p-3}^{p-2} - \ldots +
(-1)^pC_{N-1}^0\right\}-(-1)^p \nonumber\\ &=& 2^N\gamma^N_p -(-1)^p
\ena
with $\gamma^N_p \doteq \sum_{i=1}^{p+1}(-1)^{(i-1)} C_{N+p-i}^{p-(i-1)}.$
Symmetric tensors
valued in Grassmann algebras are themselves even or odd and we write, as in
section \ref{sec3.2},
$V^p(N) ={}^0V^p(N) \oplus {}^1V^p(N) $ of dimensions $e_p \doteq dim \
{}^0V^p(N) =
2^{N-1} \gamma^N_p - (-1)^p$ and $o_p \doteq dim \ {}^1V^p(N) = 2^{N-1}
\gamma^N_p.$ A simple exercise of manipulation of formal series shows that \bea
e(t) &\doteq& \sum_p e_p t^p = {2^{N-1}-(1-t)^N \over (1+t)(1-t)^N} \nonumber\\
o(t) &\doteq& \sum_p o_p t^p = {2^{N-1} \over (1+t)(1-t)^N} \ena
The ${\Z}_2$-graded Poincar\'e polynomial encodes the dimension of $V^p(N)$ and
is defined as $P(t) \doteq e(t) + \epsilon \ o(t)$ where $\epsilon$ is only a
formal symbol. This result
was already obtained in \cite{K} using K\"unneth-like formulae, with no
reference to currents in Grassmann algebras. Notice that the dimension of the
space of cyclic cocycles itself differs from $e_p$ and $o_p$ by a trivial
additive constant (the number $1$) depending upon the parity of
$p.$ Indeed $H_\lambda^p={}^0H_\lambda^p\oplus{}^1H_\lambda^p$ with $dim \
{}^0H_\lambda^p = o_p$ in all cases but $dim \ {}^0H_\lambda^p = e_p$ when $p$
is odd and $dim \ {}^0H_\lambda^p = e_p +1$ when $p$ is even.

\indent

{\sl Second method}

\indent

The other method amounts to compute directly the dimension of the space of
those symmetric tensors that satisfy the constraint $\partial \phi = 0.$ This
is a rather tedious work but it
is interesting because it is quite explicit and can be used to exhibit a basis
of cocycles. We
start from $\phi^{\mu_1\ldots\mu_p
}(\theta^1\ldots\theta^p)$ such that
\be
\sum_{\mu_1 = 1}^N \partial_{ \mu_1} \tilde \phi^{\mu_1\mu_2\ldots\mu_p } = 0,
\ \forall \mu_2\ldots\mu_p.
\ee
We then develop $\phi$ as follows
\be
\phi^{\mu_1\ldots\mu_p }(\theta^1,\ldots,\theta^N) =
\sum_{m=0}^N\sum_{\nu_1<\ldots<\nu_m}
\phi^{\mu_1\ldots\mu_p }_{\nu_1\ldots\nu_m } \ \partial_{\nu_m}
\partial_{\nu_{m-1}} \ldots \partial_{ \nu_1} (\theta^1 \theta^2 \ldots
\theta^N) \ee
where $\phi^{\mu_1\ldots\mu_p }_{\nu_1\ldots\nu_m }$ is now real (or complex)
but totally
symmetric in $\mu$'s and totally antisymmetric in $\nu$'s. The equations to be
solved read $\forall m \in 0,1,\ldots (N-1),\ \forall
\mu_1<\nu_1<\nu_2<\ldots<\nu_m,\ \forall \mu_2, \mu_3, \ldots, \mu_p $,
\be
\phi^{\mu_1\ldots\mu_p }_{\nu_1\ldots\nu_m }- \phi^{\nu_1\mu_2\ldots\mu_p
}_{\nu_2\ldots\nu_m\mu_1 }+
\phi^{\nu_2\mu_2\ldots\mu_p
}_{\nu_3\ldots\nu_m\mu_1\nu_1}-\ldots +
(-1)^m \phi^{\nu_m\mu_2\ldots\mu_p
}_{\mu_1\nu_1\ldots\nu_{m-1}} = 0
\ee

Notice that for a fix $m$ there are $C_N^{m+1}C_{N+p- 2}^{p-1}$ equations but
they are not
independent. A rather laborious analysis shows that, for each value of $0\leq
m\leq
N-1$ there are $\sum_{j=1}^p(-1)^{j+1} C_N^{m+j}\ C_{N+p- j-1}^{p-j}$
independent equations. This
number is equal (cf appendix B) to
\be
\sum_{j=1}^p(-1)^{j+1} C_N^{m+j}\ C_{N+p- j-1}^{p-j} =C_{p+m-1}^m \
C_{N+p-1}^{p+m} \label{bino}
\ee
The dimension of the space of symmetric tensors is already known to be equal to
$2^N\
C_{N+p-1}^p$. The dimension of $V^p(N)$ is therefore equal to
\be
dim \ V^p(N) =
2^N \ C_{N+p-1}^p - \sum_{m=0}^{N-1} C_{p+m-1}^m \ C_{N+p-1}^{p+m}
\ee
The sum appearing on the right hand side can be written also as (cf appendix B)
\be
(-1)^p + 2^N \sum_{m=0}^{p-1} (-1)^m\ 2^m \ C_{N+p-1}^{p-1-m}
\ee
The fact that the quantity $dim V^p(N)$ is equal to the value given previously
(first method)
relies on a curious identity (proved in appendix B) between binomial
coefficients, namely
\be
\sum_{m=0}^{p-1}(-1)^m\ 2^m \ C_{N+p-1}^{p-1-m} = - \sum_{m=0}^{p-1}(-1)^{m+p}
\ C_{N+m-1}^m \ee

\subsection{The case $p=1$}

\indent

The case $p=1$ is of particular interest. Indeed, those cocycles can be used to
construct several inequivalent cyclic cocycles on the super-circle
$C^\infty(S^1) \otimes Gr(N)$ and this, in turns, allows one to construct many
inequivalent central extensions of graded loop algebras
(i.e. super Ka\v c-Moody algebras)\cite{CFRS}. Even, in this simple case $p=1$,
the total number of
independent cyclic one-cocycles grows rapidly with $N$. Previous formulae give
$dim \ {}^0H_\lambda^1 = (N-1)2^{N-1} +1$ and $dim \ {}^1H_\lambda^1
=(N-1)2^{N-1}$. Altogether $ d = dim \ H_\lambda^1 =(N-1)2^N +1.$ In the cases
$N=1,2,3$, for instance, one gets $d =1$, $d=3+2\epsilon \approx 5$ and
$d=9+8\epsilon\approx 17$.
For example, in the case N=2, p=1, the most general current of degree 1 reads
$\phi=\phi^1 \partial_1 + \phi^2\partial_2$. Taking into account cyclic
symmetry does not bring anything new in this simple case. Let us write
$\phi^i = \phi^i_0 + \phi^i_1 \theta^1 + \phi^i_2 \theta^2 + \phi^i_{12}
\theta^1 \theta^2$. Then $\tilde \phi^i = \phi^i_0 - \phi^i_1 \theta^1 -
\phi^i_2 \theta^2 + \phi^i_{12} \theta^1 \theta^2$. The condition $\partial
\phi = 0$ reads
\beano
\partial_1 \tilde \phi^1 + \partial_2 \tilde \phi^2 &=& 0 \\ -\phi^1_1 +
\phi^1_{12}\theta^2- \phi^2_2 - \phi^2_{21}\theta^1 &=& 0 \enano
Therefore closed currents are such that $\phi^1_1+\phi^2_2=0$, $\phi_{12}^1=0$
and $\phi_{12}^2=0$. These three algebraic relations are independent, so we
have $8-3=5$ independent cocycles (closed currents). The most general one reads
\be
\phi = [x+y\theta^1+z\theta^2]\partial_1 +[x'+t\theta^1 -y\theta^2]\partial_2
\ee
Let us take, for instance, $\phi \doteq \theta^1\partial_1 - \theta^2
\partial_2$. Let $a,b\in Gr(2)$, with
$a=a_0+a_1\theta^1+a_2\theta^2+a_{12}\theta^1\theta^2$ (and a similar notation
for $b$).
The bilinear form on $Gr(2)$ corresponding to $\phi$ is $\Phi(a,b) = \langle
\phi\ ,\ a db \rangle $ but $db=(b_1+b_{12}\theta^2)d\theta^1 + (b_2 -
b_{12}\theta^1)d\theta^2$ so that $\Phi(a,b)= a_1 b_2+ a_2 b_1.$ Considering
cyclic cocycles as bilinear forms on $Gr(2)$ and using the notation introduced
in eq. (\ref{eq.eps}), one can write a basis of this space as follows.
\be
\begin{array}{c|c}
\mb{even cocycles} & \mb{odd cocycles} \\ \epsilon_{1,1} &
\epsilon_{1,12}-\epsilon_{12,1} \\ \epsilon_{2,2} &
\epsilon_{2,12}-\epsilon_{12,2} \\ \epsilon_{1,2}+\epsilon_{2,1} &
\end{array}
\ee
For instance, $\Phi \doteq \epsilon_{1,2}+\epsilon_{2,1}$ evaluated on $a,b\in
Gr(2)$, gives $\Phi(a,b)=a_1 b_2 + a_2 b_1$ as we already know. We refer to the
article \cite{CFRS} for more details concerning the case $p=1$. Since the main
subject of the present article is the study of the case $p>1$, we wish now to
illustrate the previous techniques by studying an example with $p=2, N=2$.

\newpage

\subsection{An explicit example: N=2, p=2\label{sec3.7}}

\indent

An arbitrary current of degree $2$ on
$Gr(2)$ reads $\phi = \phi^{11} \partial_1\vee \partial_1 + \phi^{12}
\partial_1\vee \partial_2+ \phi^{21} \partial_2\vee \partial_1+
\phi^{22} \partial_2\vee \partial_2.$
We recall this operator is symmetric, i.e. $\phi^{12}=\phi^{21}$.

Taking into account the symmetry, we see that \bea
\phi^{11} &=& \phi^{11}_0 + \phi^{11}_1 \theta^1 + \phi^{11}_2 \theta^2 +
\phi^{11}_{12}\theta^1\theta^2 \nonumber\\ \phi^{22} &=& \phi^{22}_0 +
\phi^{22}_1 \theta^1 + \phi^{22}_2 \theta^2 +
\phi^{22}_{12}\theta^1\theta^2 \nonumber\\ \phi^{12} = \phi^{21}
&=& \phi^{12}_0 + \phi^{12}_1 \theta^1 + \phi^{12}_2 \theta^2 +
\phi^{12}_{12}\theta^1\theta^2
\ena
Then,
\bea
\tilde \phi^{11} &=& \phi^{11}_0 - \phi^{11}_1 \theta^1 - \phi^{11}_2 \theta^2
+
\phi^{11}_{12}\theta^1\theta^2 \nonumber\\ \tilde \phi^{22} &=& \phi^{22}_0 -
\phi^{22}_1 \theta^1 - \phi^{22}_2 \theta^2 +
\phi^{22}_{12}\theta^1\theta^2 \nonumber\\ \tilde \phi^{12} = \tilde \phi^{21}
&=& \phi^{12}_0 - \phi^{12}_1 \theta^1 - \phi^{12}_2 \theta^2 +
\phi^{12}_{12}\theta^1\theta^2
\ena
The condition $\partial \phi = 0$ read
\bea
\partial_1 \tilde \phi^{11} + \partial_2 \tilde \phi^{21} &=&
- \phi^{11}_1 + \phi^{11}_{12} \theta^2
- \phi^{12}_2 - \phi^{12}_{12}\theta^1
= 0 \nonumber\\
\partial_1 \tilde \phi^{12} + \partial_2 \tilde \phi^{22} &=&
- \phi^{12}_1 + \phi^{12}_{12}\theta^2
- \phi^{22}_2 - \phi^{22}_{12} \theta^1
= 0
\ena
So we get the constraints
\bea
\phi^{11}_1 &=& - \phi^{12}_2, \ \phi^{11}_{12} = \phi^{12}_{12} = 0
\nonumber\\
\phi^{22}_2 &=& - \phi^{12}_1, \ \phi^{22}_{12} = \phi^{12}_{12} = 0
\ena
Notice that we get twice the constraint $\phi^{12}_{12} = 0.$
A general closed current will therefore depend on the $7$ independent
parameters
$\phi^{11}_0, \phi^{11}_1, \phi^{11}_2,
\phi^{22}_0, \phi^{22}_1, \phi^{22}_2,
\phi^{21}_0$.
The most general cyclic cocycle of order $2$ on $Gr(2)$ reads therefore
\bea
\Phi(a_0,a_1,a_2) = &(-1)^{\sg{ a_1}}& \int \phi^{ij} a_0 \partial_i a_1
\partial_j a_2 + k \tau(a_0,a_1,a_2) \nonumber\\ = &(-1)^{\sg{ a_1}}& \left\{
\int
(\phi^{11}_0 + \phi^{11}_1 \theta^1 +\phi^{11}_2 \theta^2)
a_0\partial_1a_1\partial_1a_2 + \right.\nonumber\\ &&\ \int
(\phi^{22}_0 + \phi^{22}_1 \theta^1 +\phi^{22}_2 \theta^2)
a_0\partial_2a_1\partial_2a_2 + \nonumber\\
&&\ \left.\int (\phi^{12}_0 - \phi^{22}_2 \theta^1 - \phi^{11}_1 \theta^2)
a_0(\partial_1a_1\partial_2a_2 +\partial_2a_1\partial_1a_2) \right\}
\nonumber\\ &+& \hs{-7}k\ \tau(a_0,a_1,a_2)
\ena
where $\tau$ is the canonical cocycle described in section \ref{sec3.4} and $k$
is an arbitrary constant. All these cocycles are non-trivial, (as cyclic
cocycles, i.e. are not cyclic coboundaries since
-- with the exception of $\tau$ -- they can be written as currents) and are
inequivalent. By
evaluating those Berezin integrals on the elements of a basis $(\theta)^I =1,
\theta^1,
\theta^2, \theta^{12} $ of $Gr(2)$ we can rephrase the above results by saying
that $H_\lambda^2 = {\C}^4 \oplus {\C}^4 $ is generated by four even generators
and four odd
generators that, when expressed as (independent) multilinear forms, with a
notation
introduced in (\ref{eq.eps}), read

\be
\begin{array}{c|c}
\mbox{ even cocycles} &\mbox{odd cocycles}\\ &\\
\tau = \epsilon_{0,0,0}
&\epsilon_{1,1,1} \\
\epsilon_{1,1,12}+\epsilon_{12,1,1}-\epsilon_{1,12,1}
&\epsilon_{2,2,2} \\
\epsilon_{2,2,12}+\epsilon_{12,2,2}-\epsilon_{2,12,2}
&\epsilon_{1,1,2}+\epsilon_{1,2,1}+\epsilon_{2,1,1}\\
\epsilon_{1,2,12}+\epsilon_{12,1,2}+\epsilon_{2,1,12} +
\epsilon_{12,2,1}-\epsilon_{1,12,2}-\epsilon_{2,12,1} &
\epsilon_{2,2,1}+\epsilon_{2,1,2}+\epsilon_{1,2,2} \end{array}
\ee
For instance, the second of the above even cocycles is a tri-linear form over
$Gr(2)$ which, when evaluated on a triple $(a,b,c)$ gives
$a_1b_1c_{12}+a_{12}b_1c_1-a_1b_{12}c_1$, with $a=a_0 +a_1\theta^1 +a_2\theta^2
+a_{12}\theta^1\theta^2$, and similar decompositions for $b$ and $c$.

The above construction using closed currents gives automatically non-trivial
cyclic cocycles. This is an important remark. Indeed, if we had tried to solve
directly (\cite{CJK}) the algebraic equations $b\Phi = 0$ and $\lambda \Phi =
\lambda$ in this case, we would have found that the
space of cocycles is actually $Z_\lambda^2 = {\C}^5 \oplus {\C}^6$ and is
generated
therefore by three more multilinear forms that cannot be written as currents.
Then, one
would have had to show that these objects generate the vector space of cyclic
coboundaries,
$B_\lambda^2 = {\C} \oplus {\C}^2$ so that $H_\lambda^2 = {\C}^2 \oplus {\C}^4$
as it should. Explicitly, one finds that coboundaries are generated by
\be
\begin{array}{ll}
\epsilon_{0,0,1} + \epsilon_{0,1,0} + \epsilon_{1,0,0} &=
b(\epsilon_{0,1}-\epsilon_{1,0}) =
(2i\pi)^{-1}S(\epsilon_1)\\
\epsilon_{0,0,2} + \epsilon_{0,2,0} + \epsilon_{2,0,0} &=
b(\epsilon_{0,2}-\epsilon_{2,0}) =
(2i\pi)^{-1}S(\epsilon_2)\\
\epsilon_{0,0,12} + \epsilon_{0,12,0} + \epsilon_{12,0,0} &=
b(\epsilon_{0,12}-\epsilon_{12,0})
= (2i\pi)^{-1}S(\epsilon_{12})
\end{array}
\ee
The meaning of the $S$ operator in the above expression is explained in
appendix A.

\vs{4}

\appendix

\indent

{\Large{\bf{Appendix}}}

\sect{The action of operators $S$ and $B$ in the Grassmannian case}

\indent

Several operators called $b$, $\lambda$, $A$, $B_0$, $B$, $S$ belong to the zoo
of non-commutative
differential geometry. The first, the Hochschild coboundary operator $b$ was
already known before \cite{AC1}. The others have been invented by \cite{AC1} to
replace and generalize the usual constructions
of usual commutative geometry, when the algebra under study is no longer
commutative. The case of Grassmann algebra is rather simple and we have been
able to study its differential properties without having to introduce the
operators $A$, $B_0$, $B$ and $S.$ However, we feel instructive to
see what is the meaning of those operators in the present context. The
cyclicity operator
$\lambda$ was used in the text. $A \doteq 1+\lambda +\lambda^2\ldots+\lambda^N
$ is the cyclic antisymmetrizer; acting on a cyclic object --in our case a
closed current on Grassmann algebra-- it does essentially nothing here, besides
multiplying by a trivial numerical factor. The operator $B_0$
is called the non-symmetrized boundary operator, it maps ${\cal C}^{p+2}$ to
${\cal C}^{p+1}$ and is defined by
\be
[B_0 \Phi](a_0,\ldots,a_p)=\Phi(1,a_0,\ldots,a_p) + (-1)^{N+\sum_{i=0}^N \sg{
a_i}}\
\Phi(a_0,\ldots,a_p,1)
\ee
The B operator is the symmetrized boundary operator; it also maps ${\cal
C}^{p+2}$ to ${\cal C}^{p+1}$ and is defined by $B \doteq A \ B_0.$ The main
interest of $B$, in general, comes from
the fact that $B^2=0$ and $Bb+Bb=0.$ Morally, $B$ should be considered -- up to
a trivial
numerical factor-- as the non-commutative generalization of the De Rham
$d$-operator on forms, or better, since we are dealing with {\sl co}-chains, as
the non-commutative generalization of the
De Rham $\partial$-operator acting on currents. This is known to be true in the
classical case
of commutative geometry, and is clearly valid also in the Grassmaniann context,
since, if we
consider a closed Hochschild cochain $\Phi$, hence representable as a current
$\phi$, we see that $[B \Phi](a_0,\ldots,a_p)$ is proportional to $\langle
\phi, da_0 \vee da_1 \vee\ldots
\vee da_p\rangle$ and is therefore associated with the current $\partial \phi.$
In other
words, if we had not worked out by hand the definition of the boundary operator
$\partial$ we could have used $B$ to define it. This explain also why we did
not have to introduce $B$ explicitly before; notice
however that action of $B$ is defined on all Hochschild cochains (all
multilinear forms on
$Gr(N)$) and not only on those that are representable as non-zero currents (the
non-trivial
Hochschild cocycles). The operator $S$, in our case, brings something new. It
is called the Connes' stabilization operator, it maps ${\sl C}^p$ to ${\sl
C}^{p+2}$ and is
defined as follows
\bea
&& {1\over 2i\pi}.[S\Phi](a_0,a_1,\ldots ,a_p,a_{p+1},a_{p+2}) \,=\,
\Phi(a_0a_1a_2,a_3,\ldots,a_{p+2})\ + \nonumber\\ &&\hs{14}
+\sum_{j=2}^{p+1}\left\{ \rule{0mm}{7mm}\ \Phi(a_0,\ldots ,a_{j-1}
a_ja_{j+1},\ldots, a_{p+2}) \,+ \right.\\ &&\hs{27}\left. +\sum_{i=0}^{j-2}
(-1)^{j-i+1} \Phi(a_0,\ldots,a_ia_{i+1},a_{i+2},\ldots,a_ja_{j+1},a_{j+2}
,\ldots,a_{p+2}) \right\}\nonumber
\ena
For any associative algebra, one shows \cite{AC1} that $S$ maps cyclic
cochains, coboundaries and
cocycles of order $p$ to cyclic cochains, coboundaries and cocycles
of order $p+2$ and, in particular $H_\lambda^p$ to $H_\lambda^{p+2}.$ Also it
maps cyclic cochains to
Hochschild coboundaries. In particular, if $\Phi \in Z_{\lambda}^p$, i.e.
$b\Phi = 0$ and
$\lambda \Phi = \Phi$ then $S\Phi \in Z_{\lambda}^{p +2}$ i.e. $bS\Phi = 0$ and
$\lambda S\Phi = S\Phi$ but also $S\Phi \in B^{p+2}$ i.e. $\exists \Psi \in
{\cal C}^{p+1}$ with $S\Phi = b\Psi.$
Notice that there is no contradiction because $\Psi$ is not cyclic in general;
in other words,
if $\Phi$ is a cyclic cocycle $S\Phi$ is always a cyclic cocycle, it may be
trivial or not as a
cyclic cocycle (as the $b$ of something cyclic) but it is always trivial as a
Hochschild cocycle
(it is always the $b$ of something). When translated in the particular case of
Grassmann
algebras, these notions read as follows. \par The first observation is that the
hierarchy of operators $\tau$ can be obtained from the first
$\tau = \epsilon_0$ by applying powers of S: \be
\epsilon_{\underbrace{\mbox{\scriptsize{0$\ldots$0}}}_{2p}}= \frac{1}{p!}\
S^p\epsilon_0
\ee

Then, if $\Phi$ is a non trivial cyclic cocycle of order $p$ on
$Gr(N)$, and if we assume that it has no component along $\tau$, it can be
considered as a
closed current. $S\Phi$ is {\sl a priori} a cyclic cocycle of order $p+2$
but is is also a Hochshild coboundary
and the associated current is zero. However, since $S\Phi$ has no component
along $\tau$ and since we have already obtained all the cyclic cocycles of
order $p+2$, this implies that $S\Phi$, is not only a
Hochschild coboundary but also a cyclic coboundary. In other words, as far as
cyclic cohomology is concerned, and with the exception of $\tau$ itself, the
hierarchy of cocycles $S^p\Phi$ is
trivial: we do not get new non-trivial cocycles in this way. These remarks
explain the
(generic) features of the example detailed in section \ref{sec3.7}. To be
complete we should mention that
the (inductive) limit of $S^p(H_\lambda(Gr(N))$ when $p\rightarrow\infty$ is
called ``periodic cyclic cohomology''. The previous analysis shows that,
although $H_\lambda(Gr(N))$ is, by no means, trivial, $H_\lambda^{per}(Gr(N))$
reduces to cyclic cohomology of complex numbers.

\sect{Some combinatorial identities}

\indent

We start by proving the equality
\be
\sum_{m=0}^{N-1} C_{p+m-1}^mC_{N+p-1}^{p+m}= (-)^p + 2^N\sum_{m=0}^{p-1} (-)^m
2^m C_{N+p-1}^{p-m-1} \label{lasol}
\ee

Starting with the l.h.s. of eq. (\ref{lasol}), one can rewrite it as \beano
\sum_{j=p}^{N+p-1} C_{j-1}^{p-1}C_{N+p-1}^{j} &=& \left\{\sum_{j=p}^{N+p-1}
C_{N+p-1}^{j}\ \frac{1}{(p-1)!} \left[(1+t)^{j-1}\right]^{(p-1)} \right\}_{t=0}
\\ &=& \left\{\left[\frac{1}{(p-1)!} \sum_{j=0}^{N+p-1} C_{N+p-1}^{j}\
(1+t)^{j-1}\right]^{(p-1)} -\left[\frac{1}{1+t}\right]^{(p-1)} \right\}_{t=0}
\\
&=& \left\{\left[\frac{1}{(p-1)!}\ \frac{1}{1+t}\ \left( 1+(1+t)
\right)^{N+p-1} \right]^{(p-1)} \right\}_{t=0} - (-)^{p-1}\\ &=&
\left\{\frac{1}{(p-1)!}\ \sum_{m=0}^{p-1} C_{p-1}^{m}\
(1+t)^{(m)}\left[(2+t)^{N+p-1}\right]^{(p-1-m)} \right\}_{t=0} + (-)^p\\ &=&
(-)^p+ \left\{\frac{1}{(p-1)!} \sum_{m=0}^{p-1} C_{p-1}^{m}\ \frac{(-)^m\
m!}{(1+t)^{m+1}}\ \frac{(N+p-1)!}{(N+m)!}\ (2+t)^{N+m} \right\}_{t=0}
\enano
the last equality being just the r.h.s. of eq. (\ref{lasol}).

\indent

Now, we prove the following identity between binomial coefficients:

\be
\sum_{m=0}^{p-1}(-1)^m\ 2^m \ C_{N+p-1}^{p-1-m} = - \sum_{m=0}^{p-1}(-1)^{m+p}
\ C_{N+m-1}^m \ee

We first show that
$$
S(N,p) = \sum_{k \leq p} C_{p+N}^k x^k y^{p-k} $$
and
$$
S'(N,p)= \sum_{k \leq p} C_{k-1+N}^k x^k (x+y)^{p-k} $$
are equal. Indeed, they obey the same recursion relation, namely $$
S(N,p)=(x+y)S(N,p-1) + C_{N+p-1}^p x^p
$$
This can be proved easily by using the addition theorem for binomial
coefficients.
Moreover, $S(N,1)=S'(N,1)$ and $S(N,0)=S'(N,0)$. Therefore \be
S(N,p) = S'(N,p) \ \ \ \forall N, \, p \label{eq.toto} \ee
Notice that it can be useful to extend this equality for negative values of $p$
and $N$ by extending the Pascal's triangle upward (i.e. imposing the addition
theorem
for the $C_r^{-s}$). By differentiating this equality enough times with respect
to $y$, it can also be seen that the above is a remote consequence of the
(Pfaff)
reflection property (relating arguments $z$ and $ \frac{-z}{1-z}$) of the Gauss
hypergeometric function.

\indent

Setting now $x=-1$ and $y=2$ in eq. (\ref{eq.toto}), we obtain the equality
that we were looking for. It does not seem to be possible to give a simple
closed form formula for the above
sums, at least with the previous choice for $x$ and $y$. It may be interesting
to
notice that a nice close formula exists when $x=y=1$ and this is worked out in
\cite{G-K-P}.

\indent

We finish with the proof of the equality (\ref{bino}). The demonstration is
quite similar to the previous one. We define \bea
S_1(p,m) &=& \sum_{j=1}^{p} (-)^{j+1} C_{N}^{m+j}C_{N+p-j-1}^{p-j} \\ S_2(p,m)
&=& C_{p+m-1}^{m}C_{N+p-1}^{p+m} \ena
(we have drop the dependence in $N$ because it plays no role here). These two
binomial coefficient polynomials obey to the same recursion relation, namely
\be
S_k(p+1,m)= C_{N}^{m+1}C_{N+p-1}^{p+m} - S_k(p,m+1) \ \ k=1,2
\ee
Moreover, it is easy to compute that
$S_k(1,m)= C_N^{m+1}$, with $k=1,2$, $\forall m$, so that
\be
S_1(p,m)=S_2(p,m),\ \ \forall m,\ p.
\ee

 \end{document}